\documentstyle[twocolumn,prl,aps]{revtex}
\newcommand{\bee}{\begin{equation}}
\newcommand{\ee}{\end{equation}}
\newcommand{\ba}{\begin{array}}
\newcommand{\ea}{\end{array}}
\newcommand{\bea}{\begin{eqnarray}}
\newcommand{\eea}{\end{eqnarray}}
\begin{document}
\title{Reply to
Comment on ``Superinstantons and the Reliability of Perturbation Theory
in Non-Abelian Models"
}
\maketitle
{

In his recent Comment \cite{DAV} on our Physical Review
Letter \cite{PS}, David accepts our computations but
challenges our conclusion that in non-Abelian models
perturbation theory (PT) is untrustworthy even at short
distances. Moreover, he claims that if one
`renormalizes' PT appropriately, one recovers the standard
answers, such as the known Callan-Symanzik $\beta$-function.
We believe that David's statements are incorrect and reflect
only a certain bias which many particle physicists have
developed after so many years of `renormalized PT'.

David's main technical observation is that with
super-instanton (s.i.) boundary conditions (b.c.), the
PT expansion of the $O(N)$
invariant $G(x)=\langle s(0)\cdot s(x)\rangle$
develops infrared (IR) divergences.
Actually David's sketchy computations show only that
at $O(1/\beta^3)$ there {\it could} exist IR
divergences, not that they actually occur. That his
general arguments could be misleading is illustrated by
the fact that they apply equally to all $O(N)$ models
and it is almost certain that $O(2)$ has no such problems

In any case, let us assume that David's claim proved to be
correct and that for $N>2$ PT did produce IR divergent
answers. Would it then follow, as he claims, that therefore
the standard PT answers represent the correct asymptotic
expansion in the infinite volume limit? The answer is an
unqualified {\it no}: since the true infinite volume limit of
$G(x)$ is known to be independent of the
b.c. used (due to the Mermin-Wagner theorem),
if PT were uniformly asymptotic in $L$, then {\it any} b.c.
should yield the same infinite volume answer. Of course,
as stated in our paper, it could happen that for some
fortuitous reason, the standard answers were the correct
ones, but at the present time there is no compelling reason
to support that belief. For instance it could be that
if in fact IR divergences do occur with
s.i.b.c., they are merely a reflection
of the fact that the infinite volume limit of $G(x)$ may not
have an asymptotic power series expansion in 1/$\beta$,
but may be something more complicated (like an expansion
in powers and powers of logarithms).

Besides arguing for the existence of these IR divergences with
s.i.b.c., David claims that `upon appropriate renormalization'
even with s.i.b.c. one recovers the usual $\beta$-function. In
our opinion, his approach is incorrect, yet representative of many
physicists' unjustified way of identifying the lattice and the so called
continuum model. The lattice model is a well defined statistical
mechanics model, defined without any reference to PT. Expectation
values like $G(x)$ are well defined functions of $L$ and
$\beta$ and one can inquire about their true asymptotic expansions
in various limits. In particular one can ask if the asymptotic
expansion of $G(x)$ for $\beta\to\infty$ is uniform in $L$ , which is
the question addressed in our paper.

Another question of special interest for the lattice model
concerns renormalization group invariants, such as
$r\equiv G(\alpha x)/G(\alpha y)$, relevant for constructing
and characterizing the continuum limit of the lattice model.
Computed on an infinite lattice,
this defines a unique function of $\beta$ and $\alpha$ and one
can study how one must adjust $\beta$ as
a function of the scale factor $\alpha$, so that $r$ can be rendered
independent of $\alpha$ for $\alpha\to\infty$; then
$d(1/\beta)/d\ln \alpha$ is
defined as the Callan-Symanzik $\beta$-function. Obviously, if at
some $\beta$ the model is in a massless phase, as we believe to be the
case for {\it all} $O(N)$ models for $\beta$ sufficiently large
\cite{P,Lat92}, then the true $\beta$-function vanishes.
On the other hand, in a massive phase, the $\beta$-function
will be non-vanishing and representative of the properties of
the critical point reached as $\alpha\to\infty$ at fixed $r$.

These considerations have nothing to do with PT nor with the question
whether the critical point is at finite $\beta$ or not. However since the
general belief, disputed only by us, is that for $N>2$
$\beta_{crt}=\infty$,
over the years it has become common
to compute the $\beta$-function perturbatively, by inserting in $r$
the PT expression for $G(x)$. Strictly speaking such a procedure is
dubious since there is no reason to believe PT for $\alpha\to\infty$, yet
it has been accepted (see for instance \cite{LWW}) because if one
inserts the PT answers with ordinary b.c. one finds the usual
$\beta$-function. As we showed in our paper, with s.i.b.c. the
answer is changed.

David claims that the reason for this finding is our failure
to `renormalize' appropriately, certain `singular' operators which
enter his computations. His discussion apparently refers to the formal
continuum limit obtained by doing PT on the lattice and letting
$L\to\infty$ in each term of the PT series.
This claim warrants two remarks:

Firstly, to really construct the continuum limit of the
nonperturbatively defined (lattice)
model in a box, one would have to consider the double limit in
which $\beta$ is driven to its critical value $\beta_{crt}$ and $L$ is
sent to $\infty$ in a correlated fashion. To obtain a nontrivial limit,
the fields have to be multiplied by numbers diverging in that
limit, a fact that is conventionally denoted as `field strength
renormalization'. Also if $\beta_{crt}=\infty$, as
David presumably believes (but we do not, as already stated above),
then $\beta$ has indeed to be sent to $\infty$, a fact that
could be interpreted as `infinite coupling constant
renormalization'. On the other hand, even if $\beta_{crt}=\infty$,
unless $L$ diverges sufficiently slowly ($L<O(\exp(2\pi\beta/(N-1))$),
the $O(N)$ invariance of the lattice model is restored, hence
fixing the spin at the origin (imposing s.i.b.c.) or not
will have no effect; therefore the renormalizations needed
to control the continuum limit of the lattice model
would be the same with Dirichlet or s.i. b.c.. Renormalizations that
are different for Dirichlet and s.i. b.c., such as David claims to find
in the formal continuum limit, can only arise in the nonperturbative
continuum limit if the volume is of vanishing physical size.

Secondly and more importantly, interesting as they may be,
these issues have nothing to do
with the problem addressed in our paper, which is concerned
with a simpler question: is the asymptotic expansion of $G(x)$
at fixed $L$, $\beta\to\infty$ uniform in $L$? For this problem
the whole discussion of short distance expansion, operator
mixing and renormalization is clearly irrelevant.
Via the s.i.b.c. example we showed that even at fixed lattice distance,
such as $x=1$, the answer to the question
is negative. We then showed that, unsurprisingly, this effect
does not disappear as one increases the (lattice) distance, hence the
PT computation of the $\beta$-function described above
is affected; note that
this determination of the $\beta$-function proceeds by analyzing
the long distance behavior of the two point function on the unit
lattice and again has nothing to do with a possible continuum limit
or renormalization. Therefore,
even if we assume that David's claim is correct and that starting at
$O(1/\beta^3)$ the lattice PT answers with s.i.b.c. are IR divergent,
there is no basis for his conclusion that therefore the accepted
$\beta$-function for the lattice $O(N)$ models is the correct one.

Finally, let us point out that irrespective of the more technical
details discussed above, David's main claim, that the answers
produced by PT with standard b.c. are the correct ones, ignores
the other important observation contained in our paper, namely
the existence of the superinstantons. As we state in the paper,
for {\it any} b.c., as $L\to\infty$, the correct PT must
include expansions around  the `superinstanton gas', since the
trivial vacuum becomes degenerate. Consequently, if David is right
and there are IR divergences with s.i.b.c., then the correct PT
performed with {\it any} b.c. can be expected to be IR divergent.

\par
\medskip\noindent
Adrian Patrascioiu\par
Physics Department, University of Arizona\par
Tucson, AZ 85721, U.S.A.\par
\medskip\noindent
Erhard Seiler\par
Max-Planck-Institut f\"ur Physik \par
-- Werner Heisenberg-Institut -- \par
F\"ohringer Ring 6 \par
80805 Munich, Germany \par
}


\begin{references}
%
\bibitem{DAV} F.David, {\it Comment on ``Superinstantons and the
Reliability of Perturbation Theory in Non-Abelian Models''},
$\langle$hep-lat 9504017$\rangle$.
%
\bibitem{PS} A.Patrascioiu and E.Seiler, Phys. Rev. Lett. {\bf 74} (1995)
1920-1923.
%
\bibitem{P}A.Patrascioiu, {\it Existence of Algebraic Decay in
 non-Abelian Ferromagnets}, University of Arizona preprint AZPH-TH/91-49.
%
\bibitem{Lat92}A.Patrascioiu and E.Seiler, {\it Percolation Theory and
the Existence of a Soft Phase in 2D Spin Models}, {\sl Nucl.Phys.B.(Proc.
Suppl.)} {\bf 30} (1993) 184.
%
\bibitem{LWW} M.L\"uscher,P.Weisz, U.Wolff, {\sl Nucl.Phys.} {\bf B359}
(1991)221.
%
%
\end{references}
\end{document}